\documentclass[aps,prl,twocolumn,groupedaddress,showpacs]{revtex4}
\usepackage{graphics}

\begin{document}


\boldmath
\title{Einstein's $R^{\hat{0 }\hat{0}}$ equation 
for nonrelativistic sources
derived from \\
Einstein's inertial motion
and the Newtonian law for relative acceleration}
\unboldmath

\author{Christoph Schmid}      
\email{chschmid@itp.phys.ethz.ch}

\affiliation{ETH Zurich, Institute for Theoretical Physics, 
8093 Zurich, Switzerland}

\date{\today}

\begin{abstract}
With Einstein's  
inertial motion (freefalling and nonrotating relative to gyroscopes),
geodesics for nonrelativistic particles can intersect repeatedly,
allowing  one to 
compute the space-time curvature  
$R^{\hat{0} \hat{0}}$ exactly.
Einstein's  
$R^{\hat{0} \hat{0}}$  
for strong gravitational fields
and for relativistic source-matter
     is {\it identical} with
the Newtonian expression
for the relative radial acceleration
of neighbouring freefalling  
test-particles, spherically averaged.~---
Einstein's field  equations follow from Newtonian experiments,
local Lorentz-covariance, and 
energy-momentum conservation combined with the Bianchi identity.
\end{abstract}
%


\maketitle


Up to now, a rigorous derivation of Einstein's field equations for general
relativity has been lacking:
Wald~\cite{Wald.heuristic} writes
``a clue is provided'',
``the correspondence suggests the field equation''.
Weinberg~\cite{Weinberg.guess} 
takes the
"weak static limit", makes a "guess", 
and argues with "number of derivatives".  
Misner, Thorne, and Wheeler~\cite{MTW.six.routes} 
give
"Six Routes to Einstein's field equations", among which they recommend 
(1)~``model geometrodynamics after electrodynamics'', 
(2)~''take the variational principle with 
only a scalar linear in second derivatives
of the metric and no higher derivatives''. 

In contrast, we give 
a rigorous derivation of Einstein's  field equations 
for general relativity.

The crucial input is
Einstein's concept of 
           inertial 
motion  (freefalling and nonrotating relative to comoving gyroscopes):
the worldlines of 
freefalling
nonrelativistic test-particles, 
geodesics,
can intersect repeatedly. 
Two stones released one after the other
from rest at the North Pole 
freefalling into a vertical
well through the center of the Earth to the South Pole 
and back through the center of the Earth to the North Pole:
Einstein's geodesics cross repeatedly,  
violating the spacetime analogue of Euclid's axiom of parallels,
evidence that space-time is curved.

In our decisive first  step,
we prove that the 
exact  space-time curvature 
encoded in 
$R^{\, \hat{0}}_{\, \, \,  \hat{0}} (P)$ 
(curvature-side of Einstein's equation) is
          fully {\it determined}
by measuring test-particles 
which are 
          {\it quasi-static}
(or non-relativistic) 
relative to an observer with worldline through $P$
and $\bar{u}_{\rm obs} (P) = \bar{e}_{\hat{0}} (P)$.
It is superfluous to re-measure 
$R^{\, \hat{0}}_{\, \, \,  \hat{0}}$ with relativistic
test-particles.~---
Further, we prove that $R^{\, \hat{0}}_{\, \, \,  \hat{0}} $ is 
        {\it identical}
with the 
        Newtonian  
expression for the 
relative radial acceleration of neighboring freefalling 
nonrelativistic test-particles,
spherically averaged,
for gravitational fields of arbitrary strength
and for arbitrary (relativistic) source-matter.~---
Hats over indices denote  Local Ortho-Normal Bases (LONB)
following Misner, Thorne, and Wheeler, and we use their sign conventions 
     \cite{MTW.six.routes}.

The {\it explicit} expression
for Einstein's curvature $R^{\, 0}_{\, \, \,  0}$
in 
     general coordinates 
in terms of Christoffel connection coefficients
$(\Gamma^{\alpha}_{\, \, \, \beta})_{\gamma} \equiv 
\Gamma^{\alpha}_{\, \, \, \beta \gamma} $
has 106 terms, utterly uninstructive. 
Newtonian relative acceleration
in general Lagrangian 3-coordinates (e.g. comoving with the wind) 
has the same number of 106 uninstructive terms.

The expressions for 
Einstein's $R^{\, \hat{0}}_{\, \, \, \hat{0}} (P)$ and
the Newtonian relative acceleration
are extremely simple and 
     {\it explicitely identical}
with the following choices:  
(1)~We work with Local Ortho-Normal Bases (LONBs) 
in Cartan's method.
(2)~We use a 
      {\it primary observer} 
(non-inertial or inertial)
with worldline through $P$,
with $\bar{u}_{\rm obs} = \bar{e}_{\hat{0}} $,
and with his spatial LONBs $\bar{e}_{\hat{i}}$ 
along his worldline.
(3)~We use the 
      primary observer's 
      spacetime slicing 
$\Sigma_t$
by radial 4-geodesics starting Lorentz-orthogonal to his worldline,
and $t \equiv$ time measured on his worldline.
(4)~Most crucial: 
for measuring 
relative accelerations of neighbouring test-particles,
we need 
     {\it auxiliary observers} 
with 
     {\it LONBs radially parallel} (at a given  time)
to the 
    primary observer's LONBs (to avoid unnecessary extra terms).
Therefore the Ricci connection coefficients for radial displacements
from the primary observer's worldline vanish,
\begin{eqnarray}
&&\quad \quad \quad \quad \quad \, \, \mbox{auxiliary observers}
\nonumber
\\
&&\mbox{with radially parallel LONBs at given time}
\nonumber
\\
&& \quad \quad \quad \quad \quad \Leftrightarrow \quad 
[(\omega^{\hat{a}}_{\, \, \,  \hat{b}})_{\hat{i}}]_{r=0} 
\, \, = \, \, 0, \,  \, \, 
\label{omega.rad.parallel LONBs}
\end{eqnarray}
%
for $(a, b, ...)=$ spacetime indices, $(i, j, ...)=$ 3-space indices.
     Eq.~(\ref{omega.rad.parallel LONBs})
is a special case of Newtonian observers 
at relative rest at a given time. 
(5)~We use 
Riemann normal 3-coordinates centered on the primary observer
on the slices of fixed time: 
$r$-coordinate lines are radial geodesics,
$r \equiv$ geodesic radial distance, 
$(\theta, \, \phi)_P \equiv$ starting angles 
of  radial geodesics to $P$, 
and $(x,  y,  z)_P$ 
with the standard connection to
spherical coordinates.

In this first paper, we treat only 
     {\it inertial primary observers}.
Hence  
the Ricci connection coefficients 
for displacements 
along the primary observer's worldine
vanish,
\begin{eqnarray}
\mbox{inertial primary observer} \,  &\Leftrightarrow& \,   
[(\omega^{\hat{a}}_{\, \, \,  \hat{b}})_{\hat{0}}]_{r = 0}
 \, \, = \, \, 0. \, \, \, 
\label{omega.inertial.primary.obs}
\end{eqnarray}
%

With
     Eqs.~(\ref{omega.rad.parallel LONBs}, 
           \ref{omega.inertial.primary.obs}),
{\it all} Ricci connection coefficients 
vanish on the worldline of the primary
inertial observer. 
Our auxiliary
observers cannot be inertial, unless spacetime is flat.

The result:
the expression for Einstein's 
$R^{\, \hat{0}}_{\, \, \,  \hat{0}}$
in terms of quasi-static (or non-relativistic) test-particles, 
for gravitational fields of arbitray strength, and 
for arbitrary (relativistic) source-matter, is 
      {\it exactly and explicitely identical}  
with the Newtonian expression,
and this expression  is 
      {\it exactly linear}
in the gravitational field,
%
\begin{eqnarray}
&&\mbox{inertial primary observer, radially parallel LONBs:} \, \, 
\nonumber 
\\
&&  
\quad \quad \quad \quad \Leftrightarrow \, \, \, \, 
R^{\, \hat{0}}_{\, \, \,  \hat{0}} 
\, \, = \,  \,
 \mbox{div} \, \vec{E}_{\rm g}
\, \,  = \, \, \mbox{div} \, \vec{g}. \, \,
\end{eqnarray}
%
In our exact operational definition in arbitrary (3+1)-spacetimes, 
the gravito-electric field
$\,  \vec{E}_{\rm g} = \vec{g} \,$
is the acceleration of
        {\it quasi-static} 
(or non-relativistic) freefalling test-particles,
measured by the chosen observer. 
But this  $\vec{E}_{\rm g}$
remains exactly valid for {\it relativistic} test-particles
in the equations of motion and in curvature calculations.

In a companion paper, we shall show in detail that for
     {\it non-inertial observers}
and for quasi-static (or non-relativistic) test-particles,
(1) the exact explicit expression for 
Einstein's $R^{\, \hat{0}}_{\, \, \,  \hat{0}}$ 
and the 19th-century Newtonian expression 
for relative acceleration of neighbouring freefalling particles,
spherically averaged, 
are
     {\it identical}, 
if one uses Einstein's 
     {\it equivalence} 
of
     {\it fictitious forces} 
and 
     {\it gravitational forces}
$\, (\vec{E}_{\rm g}, \, \vec{B}_{\rm g}), $  
which has been demonstrated explicitely in 
   \cite{CS.arXiv.July.2016},
and (2) that the two identical expressions are 
    {\it nonlinear}
in the gravitational fields.

In the  second  (trivial) step 
for deriving Einstein's $R^{\, \hat{0}}_{\, \, \, \hat{0}}$ equation,
we put non-relativistic source-matter on the matter-side of 
Einstein's equation:
it follows that Einstein's  
$R^{\, \hat{0}}_{\, \, \,  \hat{0}}$ equation 
for nonrelativistic source-matter 
and for gravitational fields of arbitrary strength
is exactly
     {\it identical} 
with the Newtonian equation
for the relative radial acceleration
of neighbouring freefalling test-particles, spherically averaged.

In a third,  well-known step, given in textbooks,
one derives the general Einstein equations 
from Einstein's $R^{\, \hat{0}}_{\, \, \, \hat{0}}$ equation 
for nonrelativistic source-matter
by using local Lorentz covariance and energy-momentum conservation 
combined with the Bianch identity.

These three steps complete 
our rigorous derivation of Einstein's field equations
for general relativity.~--- 
Additional results in
   \cite{CS.arXiv.July.2016}.

The tools needed in this paper are: 
(1)~our exact operational definition 
of the gravito-electric 
field 
$\, \vec{E}_{\rm g},$
(2)~the Ricci connection coefficients for a Lorentz boost of LONBs under a 
displacement in time,
$\, (\omega^{\hat{i}}_{\, \, \, \hat{0}})_{\hat{0}}$,
and   (3)~our identity
$E^{(\rm g)}_{\hat{i}} = - (\omega_{\hat{i} \hat{0}})_{\hat{0}}. \, $

The gravito-electric field
$\vec{E}_{\rm g}$
measured by any local observer (with his LONBs along his worldline)
is given by our exact and general operational definition 
in  arbitray (3+1)-spacetimes, 
     Eq.~(\ref{def.E}),
which is probably new.~---
In  contrast to the literature, 
we use
no perturbation theory on a background geometry, 
no weak gravitational fields.~---
$\vec{E}_{\rm g}$ is defined as the measured acceleration of 
   {\it quasistatic} freefalling test-particles 
analogous to the operational definition of the ordinary electric field,
where we  replace 
the particle's charge 
by its rest mass $m$,
%
\begin{eqnarray}
&& \quad \quad m^{-1} \, \frac{d}{dt} \, \, p_{\hat{i}} \, \, \,
\equiv \, \, E_{\hat{i}}^{\, (\rm g)} 
\nonumber
\\
&& \quad \quad \Leftrightarrow  \quad \quad \vec{a}_{\, \rm ff} 
\, \,  \,  =   \,  \, \vec{E}^{\, (\rm g)}  \, \, =  \, \, \vec{g}, 
\label{def.E}
\\
&&\mbox{for freefalling, quasistatic test-particles.}
\nonumber 
\end{eqnarray}
%
Local time-intervals  $dt$ are measured on the observer's wristwatch.
The measured 3-momentum is $p_{\, \hat{i}}$ with respect to the observer's LONB.
For a  freefalling test-particle,
     quasistatic relative 
to the observer, 
the measured gravitational 
     acceleration relative 
to the observer is
$ \, \vec{a}_{\, \rm ff}^{\, (\rm quasistatic)} = \vec{g} = \vec{E}_{\rm g}, \, $
measured by Galilei. 

The LONB-components  $\, p^{\hat{a}} \, $  are {\it directly measurable}.
This is in stark contrast to coordinate-basis components $p^{\alpha},$
which are not measurable before one has obtained $g_{\alpha \beta}$
by solving Einstein's equations for the specific problem at hand.

LONBs off the observer's worldline are not needed in
   Eq.~(\ref{def.E}),
because a particle 
released from rest (or quasistatic state)
will still be on the observer's
worldline after an infinitesimal time $\delta t$,
since $\delta s \propto (\delta t)^2 \Rightarrow 0, $
while  $\delta v \propto \delta t \neq 0$.

For a freefalling observer, 
$\vec{E}_{\rm g}  =  \vec{g} \, $ is zero on his worldline: 
Einstein's ``happiest thought of my life''.

Gravito-electric fields $\vec{E}_{\rm g}$ of arbitrary strength
can be
measured exactly with freefalling test-particles
which are {\it quasistatic} relative to the observer,
      Eq.~(\ref{def.E}).
But this same  measured $\vec{E}_{\rm g}$ is 
     {\it exactly} 
valid for 
     {\it relativistic} 
test-particles in the equations of motion.

The gravito-magnetic field  $\vec{B}_{\rm g}$
has been postulated by Heaviside in 1893 
    \cite{Heaviside}.
Our exact operational definition of 
$\vec{B}_{\rm g}$
is given in
   \cite{CS.arXiv.July.2016}.

The term ``weak gravitational fields'' for {\it local} discussions 
is often used in textbooks. 
But ``weak gravity'' is meaningless locally, 
because the gravitational field $\vec{g}$ 
and the gravitational tidal field $R^{\,\hat{0}}_{\,\ \, \hat{0}}$  are
not dimensionless.

{\it Cartan's method} with LONB-connection coefficients
is unavoidable  for our computation of curvature  
from measurements by non-inertial observers.
But Cartan's LONB method is not taught 
in almost all graduate programs
in general relativity in the USA,
and most researchers have never used Cartan's method
to solve a problem. 
Therefore we introduce  elements of Cartan's method.

{\it Ricci's LONB-connection coefficients} are illustrated 
by an  airplane on the shortest path (geodesic)
from Zurich to Chicago
and the 
Local Ortho-Normal Bases (LONBs) chosen to be in the directions 
``East'' and ``North''.
These LONBs rotate relative to the geodesic 
(relative to parallel transport) 
with a rotation angle $\delta \alpha$ per
measured path lenght $\delta s$,  
i.e. with the rotation rate $ \, \omega = (d\alpha/ds).$

For  infinitesimal displacements $\, \delta \vec{D}  \,$
in {\it any} direction, 
the rotation angle $\, \delta \alpha  \, $ of LONBs
is given by a linear map encoded by the Ricci rotation coefficients
$\, \omega_{\hat{c}},  \,$
%
\begin{eqnarray}
\delta \alpha \, &=& \, \omega_{\hat{c}} \, \, \delta D_{\hat{c}}.
\nonumber
\end{eqnarray}
The Ricci rotation coefficients are 
also called {\it connection} coefficients,
because they connect the LONBs 
at infinitesimally neighboring points by a 
rotation relative to
the infinitesimal geodesic between these points.

Cartan's LONB connection coefficients
use displacements in the {\it coordinates},
\begin{eqnarray}
\delta \alpha \, &=& \, \omega_{\gamma} \, \, \delta D^{\gamma}.
\nonumber
\end{eqnarray}

In three spatial dimensions,
the rotation of LONBs 
relative to the geodesic from $P$ to $Q$ 
must be given by 
a rotation matrix.
For a rotation in the 
$\, (\vec{e}_{\hat{x}}, \, \vec{e}_{\hat{y}})$-plane,
\[
\left(
\begin{array}
 { c}
\vec{e}_{\hat{x}}
 \\
\vec{e}_{\hat{y}}
\end{array}
\right)_Q 
\, \, = \, \, 
\left(
\begin{array}
 {*{1}{c@{\: \, \,  \:}} c}
\, \cos  \alpha & \sin  \alpha  \\
-  \sin  \alpha & \cos  \alpha
\end{array}
\right) \,
\left(
\begin{array}
 { c}
\vec{e}_{\hat{x}}\\
\vec{e}_{\hat{y}}
\label{rotation}
\end{array}
\right)_P. 
\]
%
For infinitesimal displacements, 
hence infinitesimal rotations (first derivatives in~$\alpha$), 
the rotation matrix is,
%
\[
\left(
\begin{array}
 { c}
\vec{e}_{\hat{x}}
 \\
\vec{e}_{\hat{y}}
\end{array}
\right)_Q  \, \, = \, \, 
\left[ \, 1  \, + \, \alpha \, \left(
\begin{array}
 {*{1}{c@{\: \, \,  \:}} c}
\, 0 & 1  \\
-  1 & 0
\end{array}
\right)
\right] \, 
\left(
\begin{array}
 { c}
\vec{e}_{\hat{x}}\\
\vec{e}_{\hat{y}}
\label{Lorentz.trsf}
\end{array}
\right)_P. 
\]
The infinitesimal LONB-rotation matrix  
$\, \delta R_{\hat{i} \hat{j}} \, $ is given by  the linear map
from the infinitesimal coordinate-displacement vector $D^{\hat{c}}$,
\begin{eqnarray}
\delta R_{\hat{i} \hat{j}} 
&=&    (\omega_{\hat{i} \hat{j}})_{\hat{c}} \, \, \, 
\delta D^{\hat{c}},
\nonumber
\\
  \omega_{\hat{1} \hat{2}} 
\, = \, - \, \omega_{\hat{2}  \hat{1}}
&=& \alpha_{\hat{1} \hat{2}} \, =  \, 
\mbox{rotation angle in} \, [\, \hat{1}, \, \hat{2} \, ] \, 
\mbox{plane}.  
\nonumber
\end{eqnarray}
The     $\, (\omega_{\hat{i} \hat{j}})_{\hat{c}} \, $
are the
Ricci connection coefficients.

In (1+1)-spacetime, 
the Lorentz transformation 
of the chosen LONBs 
relative to a given displacement geodesic
is a Lorentz boost
$\, L^{\hat{a}}_{\, \, \, \hat{b}}, \, $ 
\[
\left(
\begin{array}
 { c}
\bar{e}_{\hat{t}}
 \\
\bar{e}_{\hat{x}}
\end{array}
\right)_Q 
\, \, = \, \, 
\left(
\begin{array}
 {*{1}{c@{\: \, \,  \:}} c}
  \cosh  \chi & \sinh  \chi  \\
  \sinh  \chi & \cosh  \chi
\end{array}
\right) \,
\left(
\begin{array}
 { c}
\bar{e}_{\hat{t}}\\
\bar{e}_{\hat{x}}
\label{L.boost}
\end{array}
\right)_P, 
\]
with $\tanh \chi \, \equiv \, v/c,$
with $\chi$ called   ``rapidity'',
and  $\chi$  additive for successive Lorentz boosts in
the same spatial direction.  
%
For infinitesimal displacements,
the infinitesimal Lorentz boost $\, L^{\hat{a}}_{\, \, \, \hat{b}} \, $ is,
%
\[
\left(
\begin{array}
 { c}
\bar{e}_{\hat{t}}
 \\
\bar{e}_{\hat{x}}
\end{array}
\right)_Q  \, \, = \, \, 
\left[ \, 1 \, + \,  \chi \, \left(
\begin{array}
 {*{1}{c@{\: \, \,  \:}} c}
  0 & 1  \\
  1 & 0
\end{array}
\right)                                                             
\right] \, 
\left(
\begin{array}
 { c}
\bar{e}_{\hat{t}}\\
\bar{e}_{\hat{x}}
\nonumber
\end{array}
\right)_P. 
\]
%

In (3+1)-spacetime, and
with  two lower indices,
   $\, \omega_{\hat{a} \hat{b}} \,$ 
is antisymmetric for Lorentz boosts (and for rotations), 
 %
\begin{eqnarray}     
\delta L_{\hat{a}\hat{b}} \, 
= \,   (\omega_{\hat{a} \hat{b}})_{\hat{c}} \, \, \, \delta D^{\, \hat{c}},    
\, \, \,  && \, \, \,  
(\omega_{\hat{i} \hat{0}})_{\hat{c}} \, = \, - \, (\omega_{\hat{0} \hat{i}})_{\hat{c}}
\,  = \,   (\chi_{\hat{i} \hat{0}})_{\hat{c}}.
\nonumber
\end{eqnarray}
%

For a displacement in observer-time,
the {\it exact} Ricci connection coefficients 
  $ (\omega_{\hat{a}   \hat{b}})_{\hat{0}}$
of general relativity 
can be measured 
in {\it quasistatic} experiments.
But these Ricci connection coefficients 
predict the motion of {\it relativistic} particles
with the equations of motion.

Our gravito-electric field $\, \vec{E}_{\rm g} \,$
is identical with 
minus   the Ricci Lorentz-boost coefficients 
for a displacement in time,
\begin{eqnarray}
E^{(\rm g)}_{\hat{i}} 
 \, &=& \,  - \, (\omega_{\hat{i}  \,   \hat{0}})_{\hat{0}}.
\label{E.Ricci}
\end{eqnarray}
The proof:
from the point of view of the observer with his LONBs along his worldline,
the gravitational acceleration~$g_{\hat{i}} = a_{\hat{i}}^{(\rm ff \, particle)}$
of freefalling {\it quasistatic} test-particles 
(starting on the observer's worldline)
is by definition identical to  the 
{\it exact} gravitoelectric field~$E_{\hat{i}}$ of general relativity,
  Eq.~(\ref{def.E}).~---
But from the point of view of freefalling test-particles,
the acceleration of the quasistatic observer with his LONBs
is by definition identical to 
the {\it exact} Ricci LONB-boost coefficients 
   $\, (\omega_{\hat{i} \hat{0}})_{\hat{0}}$,
\begin{eqnarray}
E^{(\rm g)}_{\hat{i}} 
 \, &\equiv& \, 
[(a_{\hat{i}})_{\, \rm ff \, particle}^{\, (\rm relat.to \, obs.)}]_{\rm
             quasistatic}
\, \, = \, \, \, g_{\hat{i}}
\nonumber
\\ 
&=&  \,  - \, [(a_{\hat{i}})_{\, \rm observer}^{\, (\rm relat.to \, ff)}]_{\rm quasistatic}
\, \, \equiv \, \, - \, (\omega_{\hat{i} \hat{0}})_{\hat{0}}.
\nonumber
\end{eqnarray}
Galilei measured exact Ricci connection coefficients of general relativity:
$\, (\omega_{\hat{i} \hat{0}})_{\hat{0}} 
= \delta_{\hat{i} \hat{z}} \, (9.1 \, \mbox{m/s}^2) \, $
for LONBs in directions East, North, vertical.

Our general, exact definition of the 
gravitomagnetic field,  
$\, \vec{B}_{\rm g}/2 \equiv 
\vec{\Omega}_{\, \rm gyroscope}^{\, (\rm relat.to \, obs.)}, \,$
is discussed in 
   \cite{CS.arXiv.July.2016}.
The Ricci connection coefficients $(\omega_{\hat{i} \hat{j}})_{\hat{0}}$
equal minus the precession rate
of gyroscopes (comoving with the observer),
$(\omega_{\hat{i} \hat{j}})_{\hat{0}} = - \Omega^{\, (\rm gyro)}_{\hat{i} \hat{j}}
\equiv - \varepsilon_{\hat{i} \hat{j} \hat{k}} \Omega_{\hat{k}}^{\, (\rm gyro)}$.
These exact Ricci connection coefficients of general relativity 
were measured by Foucault in 1853.

In striking contrast, 
Christoffel  connection coefficients (for  coordinate bases), 
$\, \Gamma^{\alpha}_{\, \,  \, \beta\gamma} 
\equiv 
   (\Gamma^{\alpha}_{\, \, \,  \beta})_{\gamma},  \, $
have no direct physical-geometric meaning, 
and they cannot be known,
until the metric fields $\, g_{\mu \nu} (x) \,$
have  been obtained 
by
solving Einstein's equations for a given problem.

We write Christoffel connection 
coefficients 
with a bracket:
inside the bracket are the coordinate-basis transformation-indices 
   $\, (\alpha, \, \beta), \, $
outside the bracket is the coordinate-displacement index $\, \gamma$.

For {\it curvature} computations 
there are  two methods,
(1)~the standard method with  
coordinate bases and 
Christoffel connections  $(\Gamma^{\alpha}_{\, \, \,
  \beta})_{\gamma}$,
(2)~Cartan's method with Local Ortho-Normal Bases and LONB-connections
$(\omega^{\hat{a}}_{\, \, \, \hat{b}})_{\gamma}$.

For a primary non-inertial observer, 
     {\it Cartan's} 
method is strongly preferred,
because a radially parallel LONB-vector $\bar{e}_{\hat{0}} (P)$
off the primary observer's worldline, which is 
highly convenient for measuring relative 
radial acceleration,
does not point in the same direction as 
the natural coordinate-basis vector
$\bar{e}_0 (P) = \partial_t$ for a rotating or non-freefalling observer.

Cartan's curvature equation gives the Riemann curvature 2-form
$\, {\cal{R}}^{\hat{a}}_{\, \, \, \hat{b}} \,$ 
with 2-form components 
$\, ({\cal{R}}^{\hat{a}}_{\, \, \, \hat{b}})_{\gamma \delta} \,$  
    \cite{Cartan.curvature.formula.MTW, 
          Cartan.curvature.formula.Wald}. 
2-form components are
antisymmetric covariant components 
in a coordinate basis, denoted by Greek letters.~--- 
For an {\it inertial} primary observer
and with our {\it LONBs radially parallel,} 
{\it all} of Cartan's LONB connection coefficients
$(\omega^{\hat{a}}_{\, \, \, \hat{b}})_{\gamma}$ vanish on the worldline of
the primary observer,
   Eqs.~(\ref{omega.rad.parallel LONBs},
         \ref{omega.inertial.primary.obs}).
Therefore, in Cartan's curvature equation,
the term bilinear in the connection,  
the wedge product (antisymmetric in the suppressed 
coordinate-basis displacement-indices) 
$\, [\, \omega^{\hat{a}}_{\, \, \, \hat{c}} \wedge 
        \omega^{\hat{c}}_{\, \, \, \hat{b}} \, ] \, $
vanishes.
Hence 
Cartan's curvature 2-form $\, {\cal{R}}^{\, \hat{a}}_{\, \, \, \, \hat{b}} \,$
is equal to 
the exterior derivative $\, d \,$ of the LONB-connection 1-form 
$ \, \omega^{\, \hat{a}}_{\, \, \, \, \hat{b}} \, $ 
in notation free of form-components,
\begin{eqnarray}
{\cal{R}}^{\, \hat{a}}_{\, \, \, \, \hat{b}} \, &=& \,  
d \, \omega^{\, \hat{a}}_{\, \, \, \, \hat{b}},
\label{Cartan.curvature.component.free}
\end{eqnarray}
where $\, d \, $ denotes the antisymmetric ordinary  partial derivative,
and $(\hat{a}, \hat{b})$ are the Lorentz-transformation indices of the LONBs.

Writing explicitely 
the antisymmetric 2-form-component indices $[\mu, \nu]$ (plaquette indices)
on the left-hand-side  
and the antisymmetric pair of derivative-index and displacement-index
on the right-hand side, 
   Eq.~(\ref{Cartan.curvature.component.free}) reads,
\begin{eqnarray}
(\, {\cal{R}}^{\hat{a}}_{\, \, \, \hat{b}} \, )_{\mu \nu} &=&   
(\, d \, \omega^{\hat{a}}_{\, \, \, \hat{b}} \, )_{\mu \nu}
\,  \equiv \, 
\partial_{\mu} \,  (\omega^{\hat{a}}_{\, \, \, \hat{b}})_{\nu}
\, - \, [\, \mu \Leftrightarrow \nu \,]. \, \,  
\label{Cartan.curvature.with.components}
\end{eqnarray}
%
An instructive elementary derivation of the 
    curvature equations~(\ref{Cartan.curvature.component.free}, 
                         \ref{Cartan.curvature.with.components})
for 2-space 
is given in 
    \cite{CS.arXiv.July.2016}.

Eqs.~(\ref{Cartan.curvature.component.free}, 
      \ref{Cartan.curvature.with.components})
give our  
crucial curvature result for general relativity:
\begin{eqnarray}
&&\quad \quad \quad \, \, \, \mbox{inertial primary observer,}
\nonumber
\\
&&\quad \quad \quad \, \, \,  \, \mbox{radially parallel LONBs:} 
\nonumber
\\
&&\quad \quad \quad \quad \, \, \,  \, \, 
R^{\, \hat{0}}_{\, \, \, \hat{0}} \, \, = \, \,  
({\cal{R}}^{\, \hat{0}}_{\, \, \,\hat{i}})_{\hat{0} \, \hat{i}} \, \, =
\nonumber
\\
&&= \,  
- \, \partial_{\hat{i}} \, (\omega^{\hat{0}}_{\, \, \, \hat{i}})_{\hat{0}}
\, \,  = \, \,   \mbox{div} \, \vec{E}_{\rm g} \, \,  
= \, \,  \partial_r   <a_r^{\rm ff}>_{\rm ang.average} \, \, \, \,  
\label{crucial.result.without.source}
\end{eqnarray}
%
The last expression states that
Einstein's {\it exact}  $R^{\, \hat{0}}_{\, \, \, \hat{0}}$ curvature
is {\it identical} with
the Newtonian relative acceleration of freefalling  test-particles,
spherically averaged
for gravitational fields of arbitrary strength 
and for arbitrary source-matter (e.g. relativistic).~---
It is superfluous to re-measure  or recompute
$R^{\hat{0}}_{\, \, \, \hat{0}}$ with relativistic test-particles.

The {\it second step,} 
the derivation of Einstein's 
$R^{\, \hat{0}}_{\, \, \, \hat{0}}$ equation 
for non-relativistic sources
is now trivial:
we write the sources 
on the right-hand-side of the equation,
%
\begin{eqnarray}
&& \mbox{inertial primary observer, radially parallel LONBs,}
\nonumber
\\ 
&& \quad \quad \quad \quad \quad \quad 
\mbox{nonrelativistic sources:}
\nonumber
\\ 
&&\quad  R^{\, \hat{0}}_{\, \, \, \hat{0}}   
\quad \quad \quad \quad \quad \quad \quad \quad \quad \quad \quad 
\mbox{Einstein exact}
\nonumber
\\ 
&&\quad  =   \, \, \partial_r  \, <a_r^{\rm ff}>_{\rm ang.average} 
\quad \quad \quad \, \, \,   \mbox{Newton}
\nonumber
\\ 
&&\quad  =  \, \, \mbox{div} \, \vec{E}_{\rm g}
\quad \quad \quad \quad \quad 
\quad \quad \quad \quad \quad \,    \mbox{Gauss}
\nonumber
\\ 
&&\quad  =  \, \, - \, 4 \pi \, G_{\rm N} \,  \rho_{\rm mass}.
\label{R.zero.zero.nonrelat.sources}    
\end{eqnarray}
%

It has been often emphasized
that a fundamental difference 
between general relativity and Newtonian physics is 
the {\it non-linearity} of Einstein's equations versus 
the {\it  linearity} of the Newton-Gauss equation 
$\mbox{div} \, \vec{E}_{\rm g} = - 4 \pi G_{\rm N} \rho_{\rm mass}.$
Nothing could be farther from the truth:
We have given the proof that 
Einstein's {\it exact} $R^{\,\hat{0}}_{\, \, \, \hat{0}} (P)$  
and $\mbox{div} \, \vec{E}_{\rm g} (P)$ of Newton-Gauss
are explicitely {\it identical} and {\it linear}
in the gravitational field $\vec{g} = \vec{E}_{\rm g}$ 
for an {\it inertial  primary observer} in $P$
with $\bar{u}_{\rm obs} = \bar{e}_{\hat{0}}$,
if one uses  
our radially parallel LONBs.

But for {\it non-inertial primary observers,} 
Einstein's $R^{\hat{0}}_{\, \, \, \hat{0}}$ equation
and the Newtonian relative acceleration equation are
both   
     {\it non-linear} 
in the gravitational fields and
     {\it identical},
if one uses Einstein's {\it equivalence} 
of gravitational forces and fictitious forces 
     \cite{CS.arXiv.July.2016}. 

For a superficial reader,
Gauss's law in general relativity,
    Eq.~(\ref{R.zero.zero.nonrelat.sources}),
is ``nothing new''.
However:
(1)~Our law of
      Eq.~(\ref{R.zero.zero.nonrelat.sources})
is  derived rigorously, 
and it is 
     {\it exactly linear for inertial primary observers}. 
We have not used the usual approximation of linearized gravity.
The exact law is {\it nonlinear for non-inertial primary observers}.
(2)~Our law  of
      Eq.~(\ref{R.zero.zero.nonrelat.sources})
only holds for auxiliary observers with 
      {\it LONBs parallel along radial geodesics} 
to the LONBs of the primary observer at a given time.
(3)~Our law of
      Eq.~(\ref{R.zero.zero.nonrelat.sources})
does not hold for  the Local Inertial Frame (LIF) and the 
Local Inertial Coordinate Systems (LICS) around $P_0$ 
(used in textbooks), 
where the basis vectors are parallel 
along geodesics radiating out from one point $P_0$
in all spacetime directions.
Our law of
      Eq.~(\ref{R.zero.zero.nonrelat.sources})
cannot hold in a LIF,
because (with curvature) 
LONBs cannot be parallel on all three sides of the (geodesic) triangle:
(i)  from $P_0$ along the worldline of the primary inertial observer,
(ii) from $P_0$ along the worldline of an inertial particle  
with nonzero velocity relative to the primary observer,
(iii) from the primary to the auxiliary observer 
at a fixed time $t = t_0 + \delta t.$
(4)~Our law  of
      Eq.~(\ref{R.zero.zero.nonrelat.sources})
only holds for our {\it exact} operational definition 
of $\vec{E}_{\rm g}$ in
   Eq.~(\ref{def.E}),
which is probably new.

The {\it third step}, 
the derivation of Einstein's equations
starting from Einstein's $R^{\, \hat{0}}_{\, \, \, \hat{0}}$ equation for
nonrelativistic sources,
     Eq.~(\ref{R.zero.zero.nonrelat.sources}),
is well known and described in textbooks:
one uses local 
     {\it Lorentz covariance} and 
     {\it energy-momentum conservation} 
combined with the {\it contracted Bianchi identity}.
This completes our rigorous and simple derivation 
of Einstein's field equations of general relativity,
\begin{eqnarray}
G^{\hat{a} \hat{b}} \, &=& \, 8 \pi \, G_{\rm N} \, T^{\hat{a} \hat{b}}.
\label{Einstein.eq}
\end{eqnarray}
%








\bibliography{paperY.bib}

\end{document}